\documentclass[aps,prl,twocolumn,superscriptaddress,showpacs,bibnotes]{revtex4-1}
\usepackage{graphicx}
\usepackage{bm}
\usepackage{hyperref}
\usepackage{amsfonts}
\hypersetup{colorlinks=true,citecolor=blue,urlcolor=blue,}
\usepackage{physics}
\usepackage[shortlabels]{enumitem}
\usepackage{url}
\usepackage{float}
\usepackage{xifthen}
\usepackage{bm}
\usepackage{bbm}
\usepackage{soul}
\usepackage{tabularx}
\usepackage{multirow}
\usepackage[table,xcdraw]{xcolor}
\usepackage{parskip}

\renewcommand{\div}{\divisionsymbol}

\renewcommand{\autoref}[2][]{Fig.~\hyperref[#2]{\ref*{#2}\ifthenelse{\isempty{#1}}{}{(}#1\ifthenelse{\isempty{#1}}{}{)}}}

\begin{document}

\preprint{APS/123-QED}

\title{Highly stable aluminum air-bridges with stiffeners}

\author{Aleksey N. Bolgar}
    \email{alexgood@list.ru}
 \affiliation{
 Moscow Institute of Physics and Technology, 9
 Institutsky Lane, Dolgoprudny 141700, Russia
 }
 
\author{Daria A. Kalacheva}
 \affiliation{
 Moscow Institute of Physics and Technology, 9
 Institutsky Lane, Dolgoprudny 141700, Russia
 }
 \affiliation{
 Skolkovo Institute of Science and Technology, Skolkovo Innovation Center, Moscow 121205, Russia
 }
 \affiliation{
 National University of Science and Technology MISIS, 
 119049 Moscow, Russia
 } 

 \author{Viktor B. Lubsanov}
 \affiliation{
 Moscow Institute of Physics and Technology, 9
 Institutsky Lane, Dolgoprudny 141700, Russia
 }
 
   \author{Aleksei Yu. Dmitriev}
 \affiliation{
 Moscow Institute of Physics and Technology, 9
 Institutsky Lane, Dolgoprudny 141700, Russia
 }
  \affiliation{
 National University of Science and Technology MISIS, 
 119049 Moscow, Russia
 } 

 \author{Evgenia S. Alekseeva}
 \affiliation{
 Moscow Institute of Physics and Technology, 9
 Institutsky Lane, Dolgoprudny 141700, Russia
 }

  \author{Evgeny V. Korostylev}
 \affiliation{
 Moscow Institute of Physics and Technology, 9
 Institutsky Lane, Dolgoprudny 141700, Russia
 }
 
\author{Oleg V. Astafiev}	
\affiliation{
 Skolkovo Institute of Science and Technology, Skolkovo Innovation Center, Moscow 121205, Russia
 }
 \affiliation{
 Moscow Institute of Physics and Technology, 9
 Institutsky Lane, Dolgoprudny 141700, Russia
 }

\date{\today}

\begin{abstract}

Air-bridges play a critical role in the performance of microwave circuits integrated with superconducting quantum bits, and their mechanical stability is predominant for reliable operation. This study is devoted to the technological issues that lead to air-bridge instability. We propose an optimized bridge geometry designed to enhance mechanical resilience. Through systematic testing, we established that bridges incorporating this novel geometry achieved complete stability for lengths up to 170 micrometers in our technological processes. The findings provide an insight into the problem and a practical solution for technologists that faced with the challenges of air-bridge stability. The implementation of our technology has the potential to significantly improve the mechanical robustness of air-bridges in multi-qubit circuits  for quantum computation.

\end{abstract}


\maketitle

\section{Introduction}\label{sec:introduction}

Air-bridges play a crucial role for maintaining the continuity of lines in complex multi-qubits schemes and ensuring reliable potential equalization in grounded conductors\cite{jin2021, janzen2022, stavenga2023, chen2014, sun2022, zikiy2023high}. Such elements serve on microchips as contact connections that extend above the chip's surface, jumping over dielectric gaps to establish contact with the conductor. Without these bridges, oscillating electric fields and currents may arise between isolated segments of the conductor, leading to parasitic slot modes that can disrupt the coherence of micro-chip structures. In general terms, constructing an air-bridge usually involves two lithography processes (e-beam or optical) on two different resist layers \cite{stavenga2023, jin2018fabrication, dunsworth2018, chen2014, khalid2012, abuwasib2013fabrication, girgis2006fabrication}. The first defines the bridge's base and its deflection shape, while the second forms the bridge in the chip's plane. 

\begin{figure}[htp]
\includegraphics[width=1\linewidth]{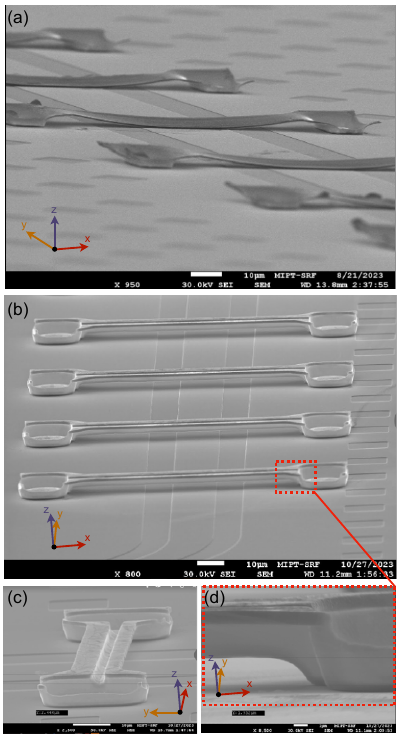}
\caption{\label{fig:bridges_comp} Mechanical instability of air-bridges and its solution: (a) SEM image of collapsed air-bridges manufactured using standard technology  with a flat profile and an aluminum film thickness of 420. (b,c) SEM image of stable air-bridges with a stiffness rib. (d) Enlarged part of the air-bridge showing the transition from the beam to the pillar.}
\end{figure}

A primary challenge in bridge fabrication arises during the final step of the technological process when the bottom resist must be removed from beneath the bridge \cite{chen2014}. In this case, the forces acting on the bridge from the liquid solvent can deform the bridge or cause it to collapse onto the substrate. This mechanical instability, as depicted in the SEM image (\autoref{fig:bridges_comp}(a)), can alter the capacitance in microwave resonators or even cause short circuits in microwave lines. Consequently, a fabrication technology that ensures the mechanical stability of air-ridges is essential for the stable operation of devices.

To remove the bottom resist, dry etching in oxygen plasma can be used \cite{khalid2012}. This method eliminates the forces that could cause bridges to collapse onto the substrate. However, in our practice, etching in oxygen plasma often leaves behind a significant amount of unetched residue on the substrate, which harmfully impacts the operation of our devices by saturating them with parasitic two-level systems (TLS)\cite{chen2014, zikiy2023high}. Moreover, for sufficiently wide air-bridges, oxygen plasma etching may fail to remove all the resist due to the shielding effect of the bridge's metallization over the electric field. An alternative liquid-free technique for bridge release utilizes HF vapor (VHF) to etch away the bottom layer of amorphous silicon oxide \cite{dunsworth2018}.This method is technically complicated and it requires some specialized equipment for VHF etching. Additionally, the HF vapors are toxic. Some groups have implemented electron-beam lithography to form masks for microbridge deposition in a lift-off process \cite{jin2021, janzen2022, stavenga2023,janzen2022aluminum, khalid2012}. However, the typical bridge size of several tens of microns does not justify the high resolution provided by electron-beam lithography. Given that, the samples investigated in this work contain hundreds of microbridges, the electron beam exposure time becomes prohibitively long. Consequently, a technology based on faster laser-beam optical lithography is more advantageous \cite{chen2014}. Typically, such technologies include a step of resist melting after its developing to achieve a smooth arched bridge shape. However, like any thermal process on qubit-containing chips, it can induce a shift in characteristics due to stimulated diffusion in the tunnel layer of Josephson junctions \cite{sun2022, hertzberg2021laser, vettoliere2020fine}. Other groups have employed grayscale optical lithography to form the bottom layer of the bridge \cite{sun2022}. This method allows to avoid the additional thermal process of resist melting.  On the other hand, grayscale lithography may not be an available option for simpler or cheap optical lithographs.

We have developed a process that utilizes straightforward optical lithography with no need for grayscale method and resist melting. \autoref{fig:bridges_comp}(b-d) displays a SEM image of an air-bridge fabricated using our technology. Bridges of this design demonstrate exceptional stability, achieving up to 100\% for lengths up to 170 microns.

This article is organized as follows. The second section analyzes the forces acting on the bridge and estimates the corresponding deformations. The third section proposes two types of bridge shape modifications which, according to our estimations, should be more stable. Finally, in the fourth section we present the fabrication results of such bridges and provide statistics on their mechanical stability. A description of the fabrication technology is included in App.\,\ref{app:fabrication}. "Crash tests" for our bridges under the influence of different airflow are shown in App.\,\ref{app:crush} 

\section{Estimating the forces acting on the air-bridge}\label{sec:estimation}

The problem of mechanical instability of microbridges can be solved by changing the shape of the bridge and modifying the fabrication process. Before presenting our solutions, it is crucial to describe the main types of forces, that act on a bridge and cause it to collapse and estimate the amount of deflection of the bridge under these forces.

\begin{figure}[htp]
\includegraphics[width=1\linewidth]{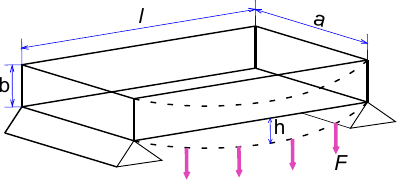}
\caption{\label{fig:model_scheme} Simplified mechanical model of a bridge is a rectangular beam supported by two pillars.}
\end{figure}

The first type of force acting on a bridge is the surface tension from the liquids used during the bridge release process (see description of the technological steps of releasing the bridge in the App.\,\ref{app:fabrication}). Various research groups use N-methyl-2-pyrollidone (NMP), isopropyl alcohol (IPA) and even acetone or water\cite{janzen2022, stavenga2023, sun2022}. The liquid surface tension coefficient $\sigma$ for these specifc liquids is 40.7, 23, 23 and 73\,$\mu$N/m, respectively. The surface tension force acting on the bridge, proportional to the total length $2l$ of its two edges. For our typical 100\,$\mu$m long bridges, we obtain forces  $F_{ST}=2l\sigma$ of 8.2, 4.6, and 14.6\,$\mu$N for NMP, IPA, and water respectively. We can estimate the amount of deflection at the bridge's center, assuming that this force is uniformly distributed over the bridge and acts normally on it. In this assessment, we consider a simplified mechanical model of the bridge as an aluminum beam in the shape of a parallelepiped, supported at both ends (see \autoref{fig:model_scheme}). For such beams, the deflection at the center under uniformly distributed load F can be expressed using the formula: 
\begin{equation}
    \Delta h = \frac{l^3F}{32Eab^3},
    \label{eq:h}
\end{equation}
where $a$,$b$,$l$ are the beam width, thickness, length respectively and $E$ is a Yong's modulus of the bridge material. For our typical bridges with the dimensions $a=20\,\mu$m, $b=0.42\,\mu$m, $l=100\,\mu$m and taking into account the surface tension forces estimated above, we obtain bridge deflection values $\Delta h$ of approximately 2.4, 1.2 and 4\,$\mu$m for NMP, IPA and water respectively. These values correspond in order of magnitude,  to the bridge height in our technology $h\approx$ 5\,$\mu$m. We can conclude that the greatest risk to our bridges comes from treating them in water, and such operation is desirable to be completely avoided. The gentliest effect is expected with IPA exposure. However, our experience of washing bridges in IPA resulted in significant amount of irremovable sediment. The cleanest results are obtained when using NMP and IPA together. Yet, even with this approach, our bridges were not robust enough. Therefore, we modified  the air-bridge's shape (see the next section) and a careful final resist washing method. The detailed fabrication process is described in App.\,\ref{app:fabrication}.

The second type of forces are hydrodynamic. During the final washing stage of the bridges, we move the sample while it is immersed in the solvent. We believe that the corresponding flow of the solvent ensures better removal of the dissolved resist particals from the bridge. The pressure force of a liquid (or gas) flow on the bridge surface can be estimated using the formula:
\begin{equation}
    F_{flow}= \frac{\rho v^2}{2}al, 
    \label{eq:flow}
\end{equation}
where $\rho$ is the liquid density and $v$ is the flow speed onto the bridge. In case of using NMP or IPA $\rho\approx$1000\,kg/$\text{m}^3$, and $v \approx0.1\div$1\,m/s for our standard sample cleaning, this leads us to estimate the hydrodynamic force in the range of 0.001$\div$0.1\,$\mu$N, which is significantly lower than the surface tension forces estimated above. Therefore, we can conclude that the rotation and oscillation of the chip when immersed in liquid do not create any risks to the mechanical stability of the bridge.

The third type of forces acting on the bridge during the fabrication process are aerodynamic forces that act on the bridge when a nitrogen flow is blown onto the chip. This is performed at the end of the bridge washing procedure to quickly remove any remaining solvent from the chip surface. This force can also be estimated using equation~\eqref{eq:flow} after substituting $\rho \approx 1$\,kg/$\text{m}^3$ and $v\approx 10\div100$\,m/s that reflects our typical blowing intensity. We estimate the aerodynamic force to be in the range of 0.1$\div$10$\,\mu$N, which is comparable to the magnitude of the surface tension forces estimated above. Thus, the nitrogen blowing process creates risks of bridge deformation similar to those caused by surface tension forces. Therefore, we conducted a series of "crash tests" on our bridges to evaluate their stability under different air flow conditions (see more in App.\,\ref{app:crush}).

\section{Modification of air-bridge shape}\label{sec:shape}

Based on our estimates for the most simplified bridge model, we can conclude that the mechanical stability of the bridge can be enhanced by increasing its thickness. To test this hypothesis, we investigate two different bridge shapes that implement this idea. The first shape increases the bridge's thickness uniformly across the surface by extending the duration of the aluminum layer deposition. In fact, doubling $b$ in eq.\,\eqref{eq:h} should reducethe deflection of the bridge by 8 times. The second shape achieves a non-uniform increase in thickness through the addition of vertical stiffener edges, without doubling the aluminum deposition duration.

\begin{figure}[htp]
\includegraphics[width=0.8\linewidth]{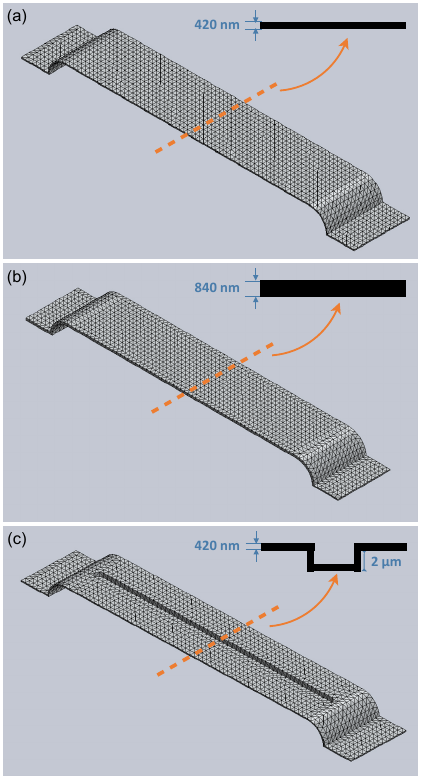}
\caption{\label{fig:sim_br} Three types of air-bridge shapes: \textbf{(a)} a standard bridge with a flat profile and an aluminum film thickness of 420\,nm; \textbf{(b)} an air-bridge with a flat profile and doubled thickness of 840\,nm;  \textbf{(c)} an air-bridge with a film thickness of 420\,nm, but enchanced with a stiffener rib 2\,$\mu m$ deep.}
\end{figure}

Thus, We investigate three modifications: our standard bridge with a flat profile and an aluminum film thickness of 420\,nm (\autoref{fig:sim_br}(a)), a bridge made of aluminum film with the doubled thickness of 840\,nm (\autoref{fig:sim_br}(b)) and a thin bridge with the film thickness of 420\,nm, but reinforced with a stiffening rib, 2 \,$\mu m$ deep (\autoref{fig:sim_br}(c)).

\begin{figure}[htp]
\includegraphics[width=1\linewidth]{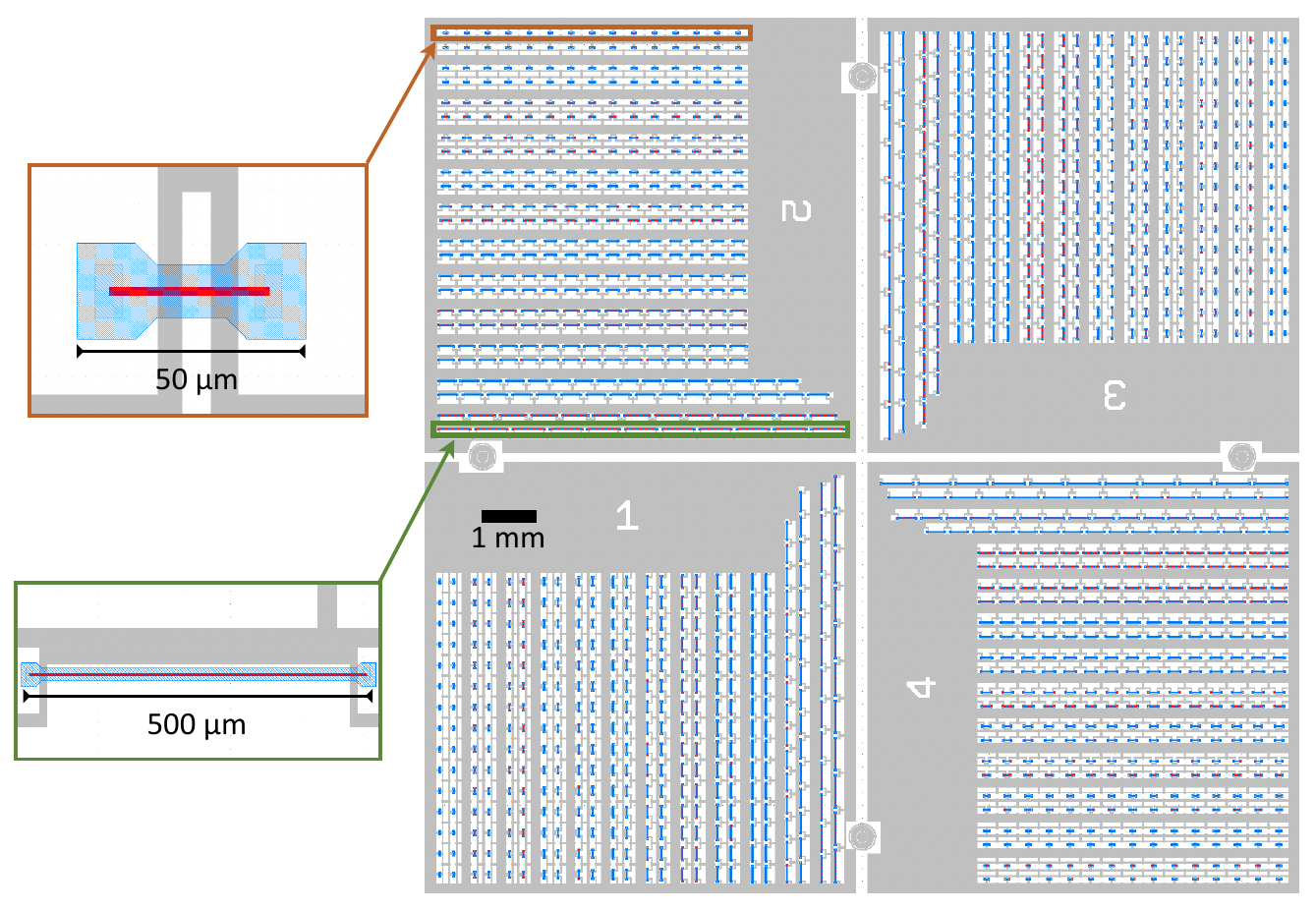}
\caption{\label{fig:design} The sample design incorporates an assortment of micro-bridge sets differentiated by orientation: horizontal bridges are delineated within a green frame, while vertical bridges are within a red frame. The test collection includes bridges of varying lengths, organized as follows: sets comprising 120 bridges each for lengths of 50, 70, 80, 100, 120, 150, 170, 190, 220, and 240 microns; sets with 64 bridges each for lengths of 270 and 300 microns; a set of 52 bridges for the length of 400 microns; and a set of 44 bridges for the length of 500 microns. Altogether, the sample presents a total of 1,440 bridges.}
\end{figure}

In order to test our predictions for improving air-bridge technology with reliable statistics, we fabricated several samples. The structure of these samples, as shown in \autoref{fig:design}, contains bridges with 14 different lengths ranging from 50 to 500\,$\mu$m. Also, to monitor the potential impact of the bridge orientation on its stability, we arranged all the bridges on the chip in two mutually perpendicular directions. The total number of bridges on the sample is 1440. Groups of bridges of the same length are connected in series to evaluate the quality of their galvanic contact. 

We investigate three different samples: the first sample is fabricated using a technology with a thin aluminum film (420\,nm) without reflowing the lower resist, but with a stiffening rib. The second one is produced with a technology that utilises a thick aluminum film (840\,nm), includes reflow of the lower resist and features a flat bridge profile. The last sample is made using a standard technology with a thin aluminum film (420\,nm), reflow of the lower resist and a flat bridge profile. 

\begin{figure}[htp]
\includegraphics[width=1\linewidth]{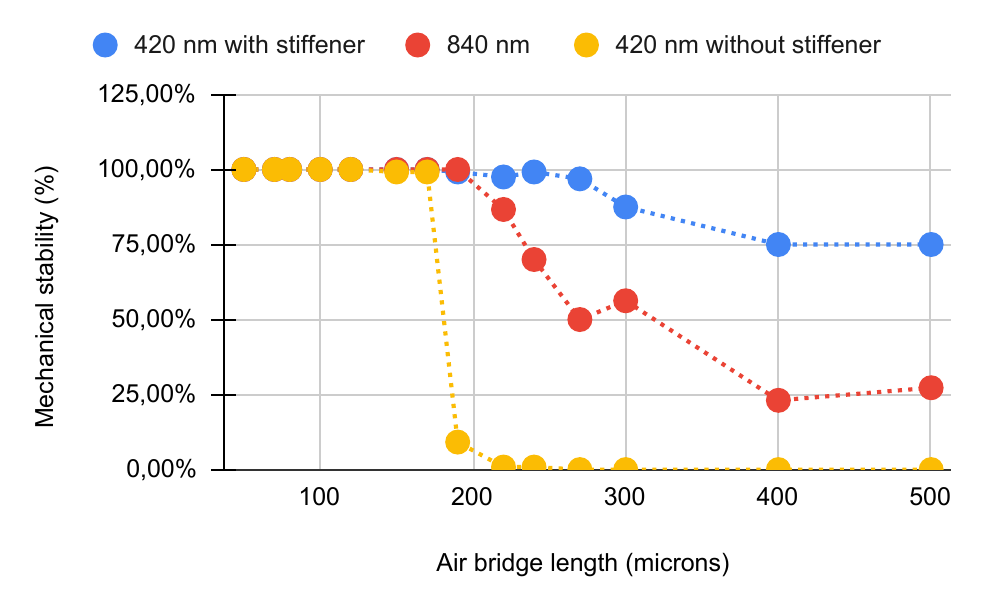}
\caption{\label{fig:chart} Comparative stability analysis of air-bridges. The chart illustrates the mechanical stability of air-bridges across a spectrum of lengths. Bridges reinforced with a stiffener exhibit the highest stability (indicated in blue), while those with a flat profile present the lowest stability levels (highlighted in yellow). Bridges composed of thick film fall in the middle range of stability (depicted in red).}
\end{figure}

For each bridge length, we tracked the mechanical stability of the bridges, quantifying it as the percentage of bridges remain unfallen. The diagram presented in \autoref{fig:chart} shows that the thin film bridges with stiffeners demonstrate absolute (100\%) mechanical stability for the length up to 170~$\mu$m based on the extensive test containing 120 bridges for each typical size. With a further increase in length, the proportion of fallen bridges gradually decreases. However, even for bridges 500\,$\mu$m long, 75\% of them remained stable. Thick film bridges also show a high level of stability up to lengths of 190\,$\mu$m, but their stability decreases more rapidly at greater lengths, with only about a quarter remaining stable at 400 and 500\,$\mu$m -- three times less than for the bridges with stiffeners. Finally, thin film bridges made using the standard technology exhibit a significant drop in stability when their length exceeds 170\,$\mu$m. Moreover, almost all these bridges collapse at lengths greater than 190\,$\mu$m.


\section{Conclusion}\label{sec:conclusion}

We analyzed the causes of air-bridge deformation and estimated the primary forces acting on the bridges during the fabrication process. Our analysis indicates that bridges predominantly deform due to two types of forces. The first type is the surface tension of the liquid during the bridge washing process. Therefore, we optimized this process to ensure that the surface tension coefficient of the solvent is minimized when it is removed from the chip surface. The second type is the aerodynamic pressure exerted on the bridges when the chip is blown with nitrogen. For bridges fabricated using our technology, we conducted a series of "crash tests" under various airflow conditions. The results of these tests enabled us to identify a set of optimal parameters for this process that enhance the mechanical stability of the bridges.

\begin{figure*}[htp]
\includegraphics[width=1\linewidth]{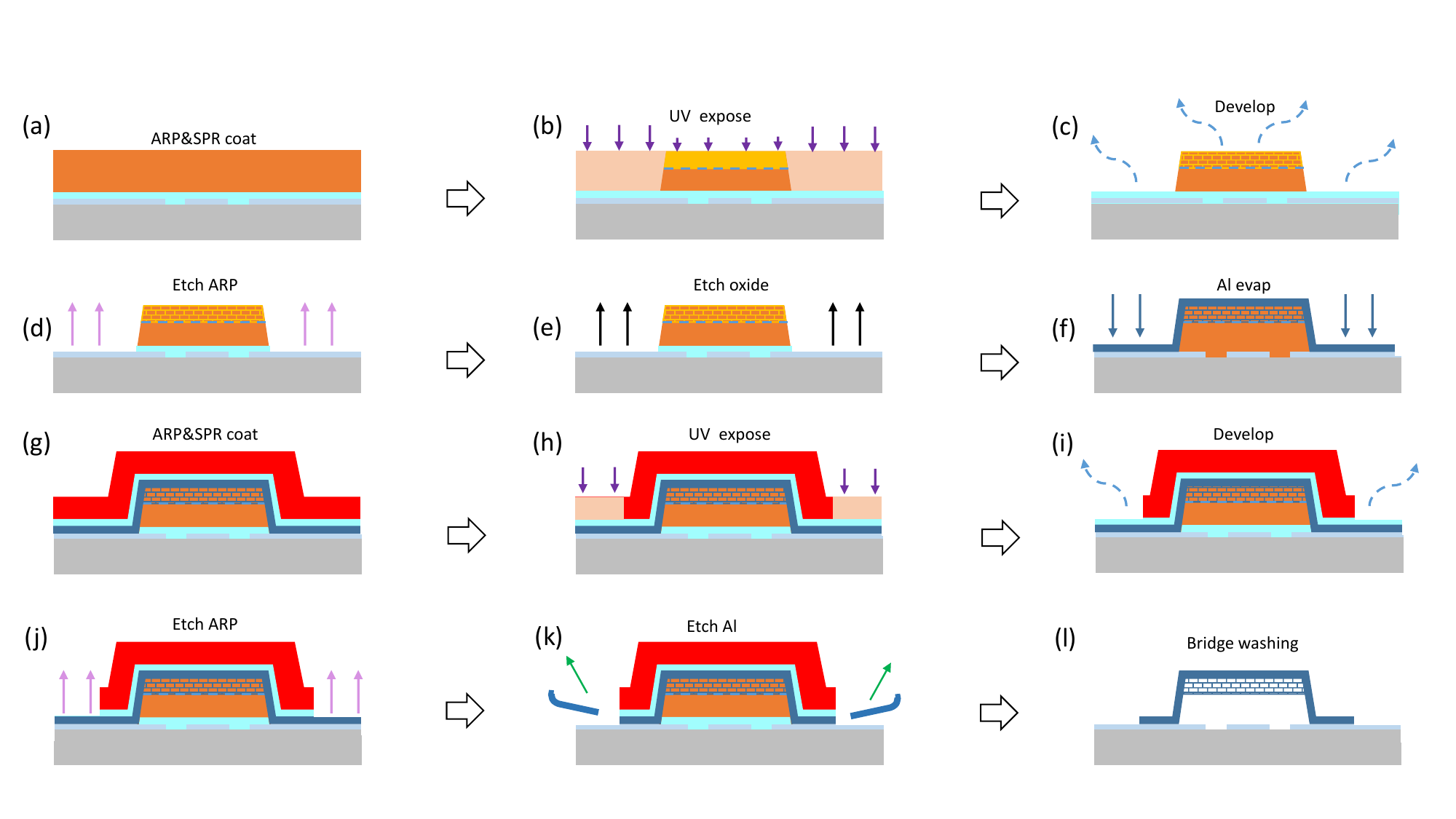}
\caption{\label{fig:steps} Schematic fabrication steps for air-bridges sample with a stiffener.}
\end{figure*}

Analytically predicting and confirming with experimental data, we demonstrated how changes in the height of bridges impact their mechanical stability. Specifically, our tests showed a significant increase in stability for bridges with a stiffening rib. We have thoroughly described the fabrication conditions for bridges of this configuration and suggest that our findings will be benefitial for technology teams facing issues with the low mechanical stability of bridges and seeking solutions to this challenge.


\section*{Data Availability}

The raw data that support the findings of this study are available on a reasonable request from the corresponding author.

\begin{acknowledgments}
The work was supported by Rosatom in the framework of the Roadmap for Quantum computing (Contract No. 868/221-D dated October 24, 2022) and Russian Science Foundation Project Grant No. 24-72-00050. This work was performed using technological equipment of MIPT Shared Facilities Center.
\end{acknowledgments}


\appendix
\section{Device fabrication}\label{app:fabrication}

Here we describe our technological process for stiffener bridges. A detailed scheme of the 12-step process is presented in \autoref{fig:steps}. The fabrication begins with (a) the coating of the substrate with an e-beam resist ARP6200.04, which is alkali-resistant and necessary for protecting the chip from etching in an alkaline developer when developing the optical resist SPR220-7, which is is coated on top. Subsequently, (b) laser lithography using a Heidelberg MLA100 system is performed for the bridge contact bases at a standard dose and for the stiffener area at a lower dose. The next step (c) involves resist development in a water solution of KOH alkaline. Then, the protective resist (d) is removed in an oxygen plasma etching process. This leads to argon etching (e) to remove aluminum oxide from the surface of the bridge contact bases, followed by (f) e-beam evaporation of an aluminum film.

The process continues with (g) applying a protective layer of ARP6200.04 resist and SPR220-7 optical resist on top, and laser lithography (h) of the bridge negative layer. Afterwards, there is another round of resist development (i) in a water solution of KOH alkaline. The protective resist is then removed again in an oxygen plasma etching process (j). The next step involves dry etching (k) of the bridge negative layer in a two-step process of $\text{BCl}_3$ and $\text{Cl}_2$ plasma respectively. Finally, the process concludes with washing out the resist (l) from under the bridge by first immersing the chip in a solvent consisting entirely of NMP, followed by the addition of a large amount of isopropyl alcohol to reduce the surface tension coefficient.


\section{Air-bridge crush test}\label{app:crush}

\begin{figure}[htp]
\includegraphics[width=0.8\linewidth]{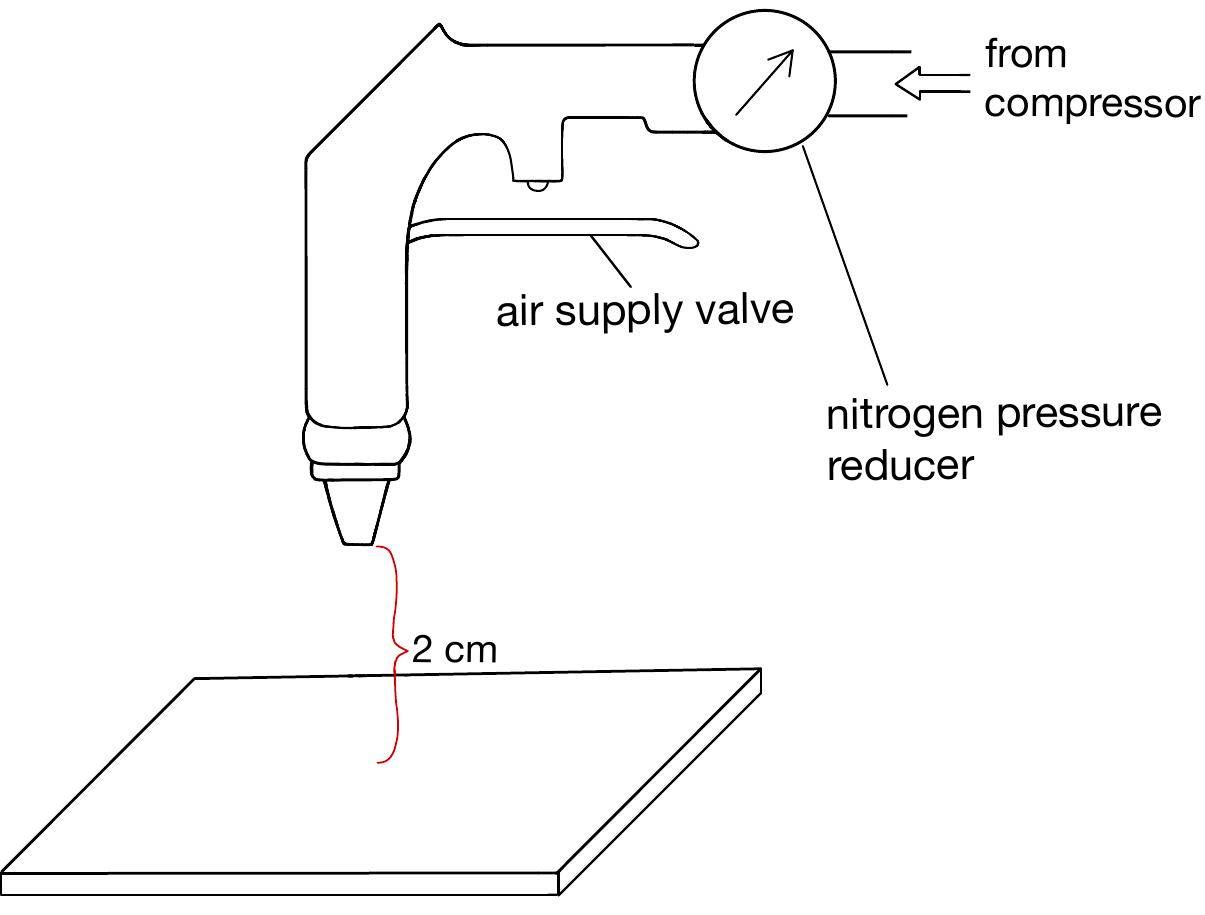}
\caption{\label{fig:crush} Air-bridge crush test setup.} 
\end{figure}

The sample was securely fixed in a horizontal position. The blow gun was positioned perpendicular to the sample at a fixed distance of 2\,cm. The pressure at which nitrogen was supplied from the compressor was regulated using the pressure reducer. The air supply valve was used to switch between uniform and pulse modes. In the uniform mode, the valve was continuously pressed, delivering air uniformly, while in the pulse mode, multiple intermittent presses were applied. The experimental setup is depicted in \autoref{fig:crush}.

Four experiments were conducted for each bridge type in the following sequence:
\begin{enumerate}
    \item Pressure of 0.5\,atm and uniform mode
    \item Pressure of 1.5\,atm and uniform mode
    \item Pressure of 1.5\,atm and impulse mode
    \item Pressure of 4.0\,atm and impulse mode
\end{enumerate}

Surface analysis of the sample was performed after each experiment to identify any bridge damage or deformation. The behavior of all three types of bridges in the above test sequence is then described. For the sample with flat bridges of 420\,nm thickness, deformation was observed in the third experiment. By the 4th experiment, some of the bridges were destroyed. Flat bridges with a thickness of 840\,nm remained undamaged in all experiments. For the bridges with a stiffener, destruction was observed only in the 4th experiment.

\begin{figure}[htp]
\includegraphics[width=1\linewidth]{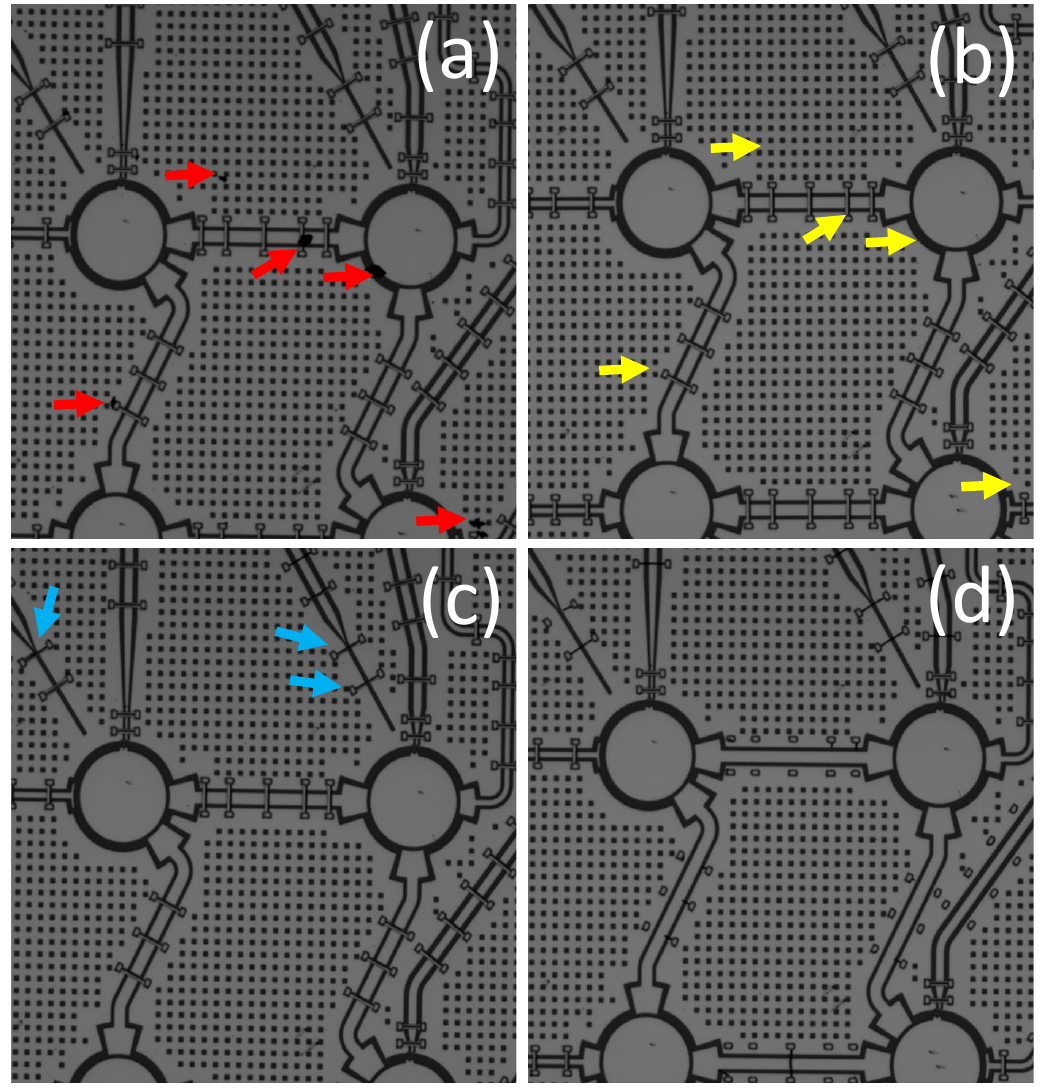}
\caption{\label{fig:crush_results} \textbf{(a)} A sample surface before the start of the experiment, red arrows indicating microparticles; \textbf{(b)} A sample surface after the experiment in the uniform mode at 1.5 atm pressure, yellow arrows indicating the location of microparticles before the start of the experiment; \textbf{(c)} A sample surface after the experiment in the impulse mode at 1.5 atm pressure, blue arrows indicate deformed bridges; \textbf{(d)} A sample surface after the experiment in the impulse mode at 4 atm pressure, most of the bridges are destroyed.} 
\end{figure}

Subsequently, a test was conducted to assess the efficiency of sample cleaning in different blowing modes. The test was performed on a sample featuring bridges with stiffening ribs. The aim was to identify the most effective mode for cleaning the sample surface without damaging the bridges. During the experiment, dust micro-particles were removed, as shown in \autoref{fig:crush_results}(a). \autoref{fig:crush_results}(b) indicates that in the uniform mode at 1.5\,atm pressure, an efficient cleaning of the sample surface occurred without damaging the bridges. \autoref{fig:crush_results}(c) illustrates that the pulse mode at 1.5\,atm pressure led to bridge deformation. The pulse mode at 4\,atm pressure led to the destruction of most bridges, as shown in \autoref{fig:crush_results}(d). Therefore, the uniform mode at 1.5\,atm pressure is optimal for cleaning the sample surface, effectively removing residual dust micro-particles without damaging the bridges.

\bibliography{manuscript}

\end{document}